\documentclass[3p,times,preprint,twocolumn]{elsarticle}
\biboptions{sort&compress}
\usepackage{enumerate}
\usepackage{mathtools}
\usepackage{color}

\usepackage[normalem]{ulem}

\makeatletter

\usepackage{amssymb,amsmath,amsthm}
\usepackage[title,toc]{appendix}
\usepackage{graphicx}
\usepackage{color}
\theoremstyle{definition}

\begin{document}

\begin{frontmatter}

\title{Doomsdays in a modified theory of gravity: A classical and a quantum approach}

\author[1,2]{Imanol Albarran}
\ead{imanol@ubi.pt}

\author[3,4]{Mariam Bouhmadi-L\'opez}
\ead{mariam.bouhmadi@ehu.eus}

\author[5,6]{Che-Yu Chen}
\ead{b97202056@gmail.com}

\author[5,6,7]{Pisin Chen}
\ead{pisinchen@phys.ntu.edu.tw}

\address[1]{Departamento de F\'{i}sica, Universidade da Beira Interior, Rua Marqu\^es D'\'Avila e Bolama,
6201-001 Covilh\~a, Portugal}
\address[2]{Centro de Matem\'atica e Aplica\c{c}\~oes da Universidade da Beira Interior (CMA-UBI), Rua Marqu\^es D'\'Avila e Bolama,
6201-001 Covilh\~a, Portugal}
\address[3]{Department of Theoretical Physics, University of the Basque Country UPV/EHU, P.O. Box 644, 48080 Bilbao, Spain}
\address[4]{IKERBASQUE, Basque Foundation for Science, 48011, Bilbao, Spain}
\address[5]{Department of Physics and Center for Theoretical Sciences, National Taiwan University, Taipei, Taiwan 10617}
\address[6]{LeCosPA, National Taiwan University, Taipei, Taiwan 10617}
\address[7] {Kavli Institute for Particle Astrophysics and Cosmology, SLAC National Accelerator Laboratory, Stanford University, Stanford, CA 94305, U.S.A.}

\begin{abstract}
By far cosmology is one of the most exciting subject to study, even more so with the current bulk of observations we have at hand. These observations might indicate different kinds of doomsdays, if dark energy follows certain patterns. Two of these doomsdays are the Little Rip (LR) and Little Sibling of the Big Rip (LSBR). In this work, aside from proving the unavoidability of the LR and LSBR in the Eddington-inspired-Born-Infeld (EiBI) scenario, we carry out a quantum analysis of the EiBI theory with a matter field, which, from a classical point of view would inevitably lead to a universe that ends with either LR or LSBR. Based on a modified Wheeler-DeWitt equation, we demonstrate that such fatal endings seems to be avoidable. 
\end{abstract}

\end{frontmatter}

\section{Introduction}
The scrutiny of extensions on General Relativity (GR) is a well motivated topic in cosmology. Some phenomena, such as the current red accelerating expansion of the universe or gravitational singularities like the big bang, would presage extensions of GR in the infra-red as well as in the ultra-violet limits. Among these extensions, the EiBI theory \cite{Banados:2010ix}, which is constructed on a Palatini formalism, is an appealing model in the sense that it is inspired by the Born-Infeld electrodynamics \cite{Born:1934gh} and the big bang singularity can be removed through a regular stage with a finite physical curvature \cite{Banados:2010ix}. Various important issues of the EiBI theory have been addressed such as cosmological solutions \cite{Avelino:2012ue,Scargill:2012kg,Bouhmadi-Lopez:2013lha,Bouhmadi-Lopez:2014jfa,Bouhmadi-Lopez:2014tna,Delsate:2012ky,Cho:2013pea}, compact objects \cite{Pani:2011mg,Pani:2012qb,Harko:2013wka,Sham:2013cya,Wei:2014dka,Olmo:2013gqa}, cosmological perturbations \cite{EscamillaRivera:2012vz,Yang:2013hsa,Du:2014jka}, parameter constraints \cite{Casanellas:2011kf,Avelino:2012ge,Avelino:2012qe}, and the quantization of the theory \cite{Bouhmadi-Lopez:2016dcf,Arroja:2016ffm}. However, some possible drawbacks of the theory were discovered in Ref.~\cite{Pani:2012qd}. Finally, some interesting generalizations of the theory were proposed in Refs.~\cite{Makarenko:2014lxa,Odintsov:2014yaa,Jimenez:2014fla,Chen:2015eha}.

As is known, the cause of the late time accelerating expansion of the universe can be resorted to phantom dark energy, which violates the null energy condition (at least from a phenomenological point of view) while remains  consistent with observations so far. Nonetheless, the phantom energy may induce more cosmological singularities in GR (curvature singularities). In particular there are three kinds of behaviors intrinsic to phantom models, which can be characterized by the behaviors of the scale factor $a$, the Hubble rate $H=\dot{a}/a$, and its cosmic derivatives $\dot{H}$ near the singular points: (a) The big rip singularity (BR) happens at a finite cosmic time $t$ when $a\rightarrow\infty$, $H\rightarrow\infty$, and $\dot{H}\rightarrow\infty$ \cite{Starobinsky:1999yw,Caldwell:2003vq, Caldwell:1999ew,Carroll:2003st,Chimento:2003qy,Dabrowski:2003jm, GonzalezDiaz:2003rf,GonzalezDiaz:2004vq,BouhmadiLopez:2009jk,Albarran:2015tga}, (b) the LR happens at $t\rightarrow\infty$ when $a\rightarrow\infty$, $H\rightarrow\infty$ and $\dot{H}\rightarrow\infty$ \cite{Nojiri:2005sx,Nojiri:2005sr,
Ruzmaikina,Stefancic:2004kb,BouhmadiLopez:2005gk,Frampton:2011sp,
Bouhmadi-Lopez:2013nma,Brevik:2011mm,Albarran:2016ewi}, (c) the LSBR happens at $t\rightarrow\infty$ when $a\rightarrow\infty$, $H\rightarrow\infty$, while $\dot{H}\rightarrow\textrm{constant}$ \cite{Bouhmadi-Lopez:2014cca,Morais:2016bev,Albarran:2015cda}.
All these three scenarios would lead to the universe to rip itself as all the structures in the universe would be destroyed no matter what kind of binding energy is involved.

Interestingly, even though the EiBI theory can cure the big bang, in Refs.~\cite{Bouhmadi-Lopez:2013lha,Bouhmadi-Lopez:2014jfa}  it was found that the BR and LR are unavoidable in the EiBI setup, hinting that the EiBI theory is still not complete and some quantum treatments near these singular events may be necessary. In this paper, we will extend the investigations in Ref.~\cite{Bouhmadi-Lopez:2016dcf} where we showed that the BR in the EiBI phantom model is expected to be cured in the context of quantum geometrodynamics. We will carry an analysis to encompass the rest of truly phantom dark energy abrupt events; i.e. the LR and LSBR.

\section{The EiBI model: The LR and LSBR}
The EiBI action proposed in \cite{Banados:2010ix} is (from now on, we assume $8\pi G=c=1$) 
\begin{equation}
\mathcal{S}_{EiBI}=\frac{2}{\kappa}\int d^4x\Big[\sqrt{|g_{\mu\nu}+\kappa R_{\mu\nu}(\Gamma)|}-\lambda\sqrt{-g}\Big]+S_m(g),\label{action1}
\end{equation}
where $|g_{\mu\nu}+\kappa R_{\mu\nu}|$ is the determinant of the tensor $g_{\mu\nu}+\kappa R_{\mu\nu}$. The parameter $\kappa$, which characterizes the theory, is assumed to be positive to avoid the imaginary effective sound speed instabilities usually associated with a negative $\kappa$ \cite{Avelino:2012ge} and $\lambda$ is related to the effective cosmological constant. $S_m$ is the matter Lagrangian. The field equations are obtained by varying \eqref{action1} with respect to $g_{\mu\nu}$ and the connection $\Gamma$. In a flat, homogeneous and isotropic (FLRW) universe filled with a perfect fluid whose energy density and pressure are $\rho$ and $p$, respectively, the Friedmann equations of the physical metric $g_{\mu\nu}$ and of the auxiliary metric compatible with $\Gamma$ are \cite{Bouhmadi-Lopez:2014jfa}
\begin{align}
\kappa H^2=&\frac{8}{3}\Big[\bar\rho+3\bar p-2+2\sqrt{(1+\bar\rho)(1-\bar p)^3}\Big]\nonumber\\
&\times\frac{(1+\bar\rho)(1-\bar p)^2}{[(1-\bar p)(4+\bar\rho-3\bar p)+3\frac{d\bar p}{d\bar\rho}(1+\bar\rho)(\bar\rho+\bar p)]^2},\label{fg}
\end{align}
and 
\begin{equation}
\kappa H_q^2=\kappa\Big(\frac{1}{b}\frac{db}{d\tilde t}\Big)^2=\frac{1}{3}+\frac{\bar\rho+3\bar p-2}{6\sqrt{(1+\bar\rho)(1-\bar p)^3}},\label{fq}
\end{equation}
where $\bar\rho\equiv\kappa\rho$ and $\bar p\equiv\kappa p$ \footnote{Notice that we are dealing with Palatini type of models which are also known as affine models. On these types of theories (c.f. the action \eqref{action1}) there is a metric $g_{\mu\nu}$ and a connection $\Gamma$ which does not correspond to the Christoffel symbols of the metric. However, it is always possible to define a metric compatible with that connection \cite{WALD} and this is the metric that we are referring to as the auxiliary metric. The same applies to the action \eqref{action2} where we denote the auxiliary metric as $q_{\mu\nu}$ and the physical metric $g_{\mu\nu}$. This is the standard and usual nomenclature in Palatini/affine theories.}. On the above equations $a$ and $b$ are the scale factor of the physical and auxiliary metrics, respectively. $\tilde t$ is a rescaled time such that the auxiliary metric can be written in a FLRW form. 

In GR, the LR and LSBR can be driven (separately) by two phantom energy models as follows \cite{Frampton:2011sp,Bouhmadi-Lopez:2014cca} 
\begin{equation}
\begin{split}
p_{LR}=-\rho_{LR}-A_{LR}\sqrt{\rho_{LR}}\,,
\qquad
p_{LSBR}=-\rho_{LSBR}-A_{LSBR}\,,\nonumber
\end{split}
\end{equation}
where $A_{LR}$ and $A_{LSBR}$ are positive constants. Therefore,
\begin{eqnarray}
\frac{\rho_{LR}}{\rho_0}&=&\Big(\frac{3A_{LR}}{2\sqrt{\rho_0}}\ln(a/a_0)+1\Big)^2,\nonumber\\
\rho_{LSBR}&=&3A_{LSBR}\ln(a/a_0)+\rho_0\,,
\label{rhophantom}
\end{eqnarray}
where we take $\rho_{LR}=\rho_{LSBR}=\rho_0$ when $a=a_0$ \cite{Frampton:2011sp,Bouhmadi-Lopez:2014cca}. The abrupt events happen at an infinite future where $a$ and $\rho$ diverge. Inserting these phantom energy contents into the EiBI model, i.e., Eqs.~\eqref{fg} and \eqref{fq}, and considering the large $a$ limit (for $\rho$ given in Eqs.~\eqref{rhophantom}), we have 
\begin{equation}
\kappa H^2\approx\frac{\bar\rho}{3}\rightarrow\infty\,,
\qquad
\kappa H_q^2\approx\frac{1}{3}\,,
\end{equation}
and
\begin{equation}
\dot{H}\approx\begin{dcases}
\frac{A_{LR}}{2}\sqrt{\rho_{LR}}\,, & \mbox{LR} \\
\frac{A_{LSBR}}{2}\,, & \mbox{LSBR} 
\end{dcases}
\end{equation}
for these two phantom energy models. Therefore, the LR and LSBR of the physical metric are unavoidable within the EiBI model while the auxiliary metric behaves as a de-Sitter phase at late time.

\section{The EiBI quantum geometrodynamics: The LR and LSBR minisuperspace model}
The deduction of the WDW equation of the EiBI model is based on the construction of a classical Hamiltonian that is promoted to a quantum operator. As shown in \cite{Bouhmadi-Lopez:2016dcf}, this can be achieved more straightforwardly by considering an alternative action which is dynamically equivalent to the EiBI action \eqref{action1}:
\begin{equation}
\mathcal{S}_a=\lambda\int d^4x\sqrt{-q}\Big[R(q)-\frac{2\lambda}{\kappa}+\frac{1}{\kappa}\Big(q^{\alpha\beta}g_{\alpha\beta}-2\sqrt{\frac{g}{q}}\Big)\Big]+S_m(g).\label{action2}
\end{equation}
In Ref.~\cite{Delsate:2012ky} it has been shown that the field equations obtained by varying the action \eqref{action2} with respect to $g_{\mu\nu}$ and the auxiliary metric $q_{\mu\nu}$ are the same to those derived from the action \eqref{action1}. Starting from action \eqref{action2} and inserting the FLRW ansatz, the Lagrangian of this model in which matter field is described by a perfect fluid can be written as (see Ref.~\cite{Bouhmadi-Lopez:2016dcf})
\begin{equation}
\mathcal{L}=\lambda Mb^3\Big[-\frac{6\dot{b}^2}{M^2b^2}-\frac{2\lambda}{\kappa}+\frac{1}{\kappa}(X^2+3Y^2-2XY^3)\Big]-2\rho Mb^3XY^3,
\end{equation}
where $X\equiv N/M$ and $Y\equiv a/b$. $N$ and $M$ are the lapse functions of $g_{\mu\nu}$ and $q_{\mu\nu}$, respectively. Note that $\rho$ is a function of $a$, i.e., $\rho=\rho(bY)$ and it is given in Eqs.~\eqref{rhophantom}.

\subsection{The classical analysis of the Hamiltonian system}
The system described by the Lagrangian $\mathcal{L}$ is a constrained system. The conjugate momenta can be obtained as follows:
\begin{align}
p_b&\equiv\frac{\partial\mathcal{L}}{\partial\dot{b}}=-\frac{12\lambda b\dot{b}}{M},\\
p_X&\equiv\frac{\partial\mathcal{L}}{\partial\dot{X}}=0,\\
p_Y&\equiv\frac{\partial\mathcal{L}}{\partial\dot{Y}}=0,\\
p_M&\equiv\frac{\partial\mathcal{L}}{\partial\dot{M}}=0.
\end{align}
Therefore, the system has three primary constraints \cite{Henneaux,Diraclecture}:
\begin{align}
p_X&\sim0,\\
p_Y&\sim0,\\
p_M&\sim0,
\end{align}
where $\sim$ denotes the weak equality, i.e., equality on the constraint surface. The total Hamiltonian of the system can be defined by \cite{Henneaux,Diraclecture}
\begin{equation}
\mathcal{H}_T=\dot{b}p_b-\mathcal{L}+\lambda_Xp_X+\lambda_Yp_Y+\lambda_Mp_M,
\end{equation}
where $\lambda_X$, $\lambda_Y$, and $\lambda_M$ are Lagrangian multipliers associated with each primary constraint. According to the consistent conditions of each primary constraint, i.e., their conservation in time: $[p_X,\mathcal{H}_T]\sim0$, $[p_Y,\mathcal{H}_T]\sim0$, and $[p_M,\mathcal{H}_T]\sim0$, one further obtains three secondary constraints \footnote{We remind that the Poisson bracket is defined as
\begin{equation}
[F,G]=\frac{\partial F}{\partial q_i}\frac{\partial G}{\partial p_i}-\frac{\partial F}{\partial p_i}\frac{\partial G}{\partial q_i},\nonumber\\
\end{equation}
where $q_i$ are the variables and $p_i$ their conjugate momenta. Notice that the repeating suffices denote the summation.
}
\cite{Henneaux,Diraclecture}:
\begin{align}
C_X\equiv&\,\lambda X-Y^3(\lambda+\kappa\rho)\sim0,\\
C_Y\equiv&\,3\lambda-3XY(\lambda+\kappa\rho)-XY^2b\kappa\rho'\sim0,\\
C_M\equiv&\,\frac{p_b^2}{24\lambda b}-\frac{2\lambda^2b^3}{\kappa}+\frac{\lambda}{\kappa}b^3X^2+\frac{3\lambda}{\kappa}b^3Y^2\nonumber\\
&-\frac{2XY^3b^3}{\kappa}(\lambda+\kappa\rho)\sim0.
\end{align}
The prime denotes the derivative with respect to $a=bY$. Furthermore, it can be shown that the total Hamiltonian is a constraint of the system:
\begin{equation}
\mathcal{H}_T=-MC_M+\lambda_Xp_X+\lambda_Yp_Y+\lambda_Mp_M\sim 0\label{htttt}.
\end{equation}
Because the Poisson brackets of the total Hamiltonian with all the constraints should vanish weakly by definition, $\mathcal{H}_T$ is a first class constraint and we will use it to construct the modified WDW equation.

This system has six independent constraints: $p_X$, $p_Y$, $p_M$, $C_X$, $C_Y$, and $C_M$. After calculating their Poisson brackets with each other, we find that except for $p_M$, which is a first class constraint, the other five constraints are second class \cite{Henneaux,Diraclecture}. The existence of the first class constraint $p_{M}$ implies a gauge degree of freedom in the system and one can add a gauge fixing condition into the system to make the constraint second class. An appropriate choice of the gauge fixing condition is $M=\textrm{constant}$ and after fixing the gauge, the conservation in time of this gauge fixing condition, i.e., $[M,\mathcal{H}_T]=0$, implies $\lambda_M=0$.

\subsection{Quantization of the system}
To construct the WDW equation, we impose the first class constraint $\mathcal{H}_T$ as a restriction on the Hilbert space where the wave function of the universe $\left|\Psi\right\rangle$ is defined, $\hat{\mathcal{H}}_T\left|\Psi\right\rangle=0$. The hat denotes the operator. The remaining constraints $\chi_i=\{M,\, p_M,\, p_X,\, p_Y,\, C_X,\, C_Y\}$ are all second class and we need to consider the Dirac brackets to construct the commutation relations and promote the phase space functions to operators \cite{Diraclecture}. Note that $C_M$ can be used to construct the first class constraint $\mathcal{H}_T$, i.e., Eq.~\eqref{htttt}, so it is excluded from the set $\chi_i$. 

The Dirac bracket of two phase space functions $F$ and $G$ are defined by \cite{Diraclecture}
\begin{equation}
[F,G]_D\equiv[F,G]-[F,\chi_i]\Delta_{ij}[\chi_j,G],
\end{equation}
where $\Delta_{ij}$ is the matrix satisfying
\begin{equation}
\Delta_{ij}[\chi_j,\chi_k]=\delta_{ik}.
\end{equation}
The existence of the matrix $\Delta_{ij}$ is proven in Dirac's lecture \cite{Diraclecture}.

According to Ref.~\cite{Diraclecture}, the second class constraints can be treated as zero operators after promoting them to quantum operators as long as the Dirac brackets are used to construct the commutation relations:
\begin{equation}
[\hat{F},\hat{G}]=i\hbar[F,G]_{D,\ (F=\hat{F},\,G=\hat{G})}.
\label{commute}
\end{equation}
This is due to the fact that the Dirac brackets of the constraints $\chi_i$ with any phase space function vanish strongly (they vanish without inserting any constraint). After some calculations, the Dirac brackets between the fundamental variables take the forms
\begin{align}
[b,p_b]_D&=[b,p_b]=1,\nonumber\\
[b,X]_D&=0,\nonumber\\
[b,Y]_D&=0,\nonumber\\
[X,Y]_D&=0,\nonumber\\
[X,p_b]_D&=f_1(X,Y,b)=f_1(b),\nonumber\\
[Y,p_b]_D&=f_2(X,Y,b)=f_2(b),
\label{Dirac}
\end{align}
where $f_1$ and $f_2$ are two non-vanishing functions. Notice that $f_1$ and $f_2$ can be written as functions of $b$ because it is legitimate to insert the constraints $C_X$ and $C_Y$ to replace $X$ and $Y$ with $b$ when calculating the Dirac brackets.

On the $XYb$ basis, if we define
\begin{align}
\langle  XYb|\hat{b}|\Psi\rangle=&b\langle XYb|\Psi\rangle,\nonumber\\
\langle XYb|\hat{X}|\Psi\rangle=&X\langle XYb|\Psi\rangle,\nonumber\\
\langle XYb|\hat{Y}|\Psi\rangle=&Y\langle XYb|\Psi\rangle,\nonumber\\
\langle XYb|\hat{p_b}|\Psi\rangle=&-i\hbar\frac{\partial}{\partial b}\langle XYb|\Psi\rangle\nonumber\\
&-f_1\frac{\partial}{\partial X}\langle XYb|\Psi\rangle-f_2\frac{\partial}{\partial Y}\langle XYb|\Psi\rangle,
\end{align}
it can be shown that the resulting commutation relations satisfy Eqs.~\eqref{commute} and \eqref{Dirac}. Furthermore, the momentum operator $\hat{p_b}$ can be written as
\begin{equation}
\langle \xi\zeta b|\hat{p_b}|\Psi\rangle=-i\hbar\frac{\partial}{\partial b}\langle \xi\zeta b|\Psi\rangle,
\label{pb2}
\end{equation}
after an appropriate redefinition of the wave functions: $\langle XYb|\rightarrow\langle \xi(X,Y,b), \zeta(X,Y,b), b|$. Therefore, in the new $\xi\zeta b$ basis, the modified WDW equation $\langle \xi\zeta b|\hat{\mathcal{H}}_T|\Psi\rangle=0$ can be written as
\begin{equation}
\frac{-1}{24\lambda}\langle \xi\zeta b|\frac{\hat{p_b}^2}{b}|\Psi\rangle+V(b)\langle \xi\zeta b|\Psi\rangle=0,
\label{wdwtotal}
\end{equation}
where the term containing $\hat{p_b}^2$ is determined by Eq.~\eqref{pb2} and its explicit form depends on the factor orderings. Note that the eigenvalues $X$ and $Y$ can be written as functions of $b$ according to the constraints $C_X$ and $C_Y$, hence it leads to the potential $V(b)$ as follows  
\begin{equation}
V(b)=\frac{2\lambda^2b^3}{\kappa}+\frac{\lambda}{\kappa}b^3X(b)^2-\frac{3\lambda}{\kappa}b^3Y(b)^2.
\end{equation}

\subsection{Wheeler-DeWitt equation: factor ordering 1 \label{3.2}}
In order to prove that our results are independent of the factor ordering, we make two choices of it. First, we consider $\langle \xi\zeta b|b^3\hat{\mathcal{H}}_T|\Psi\rangle=0$ and choose the following factor ordering:
\begin{equation}
b^2\hat{p_b}^2=-\hbar^2\Big(b\frac{\partial}{\partial b}\Big)\Big(b\frac{\partial}{\partial b}\Big)=-\hbar^2\Big(\frac{\partial}{\partial x}\Big)\Big(\frac{\partial}{\partial x}\Big),
\end{equation}
where $x=\ln(\sqrt{\lambda}b)$. Near the LR singular event, the energy density $\rho$ behaves as $\rho\propto(\ln{a})^2$. On that regime, the dependence between the auxiliary scale factor $b$ and $a$ is $b\propto a\ln{a}$. On the other hand, near the LSBR event the energy density behaves as $\rho\propto\ln{a}$ and $b$ behaves as $b\propto a\sqrt{\ln{a}}$. For both cases, the WDW equation can be written as
\begin{equation}\label{difeq1}
\Big(\frac{d^2}{dx^2}+\frac{48}{\kappa\hbar^2}\textrm{e}^{6x}\Big)\Psi(x)=0,
\end{equation}
when $x$ and $a$ go to infinity. Note that we have replaced the partial derivatives with ordinary derivatives and $\Psi(x)\equiv\langle \xi\zeta b|\Psi\rangle$. The wave function reads \cite{mathhandbook}
\begin{equation}
\Psi(x)=C_1J_0(A_1\textrm{e}^{3x})+C_2Y_0(A_1\textrm{e}^{3x}),
\end{equation}
and consequently when $x\rightarrow\infty$, its asymptotic behavior reads \cite{mathhandbook}
\begin{equation}\label{approxPsix}
\Psi(x)\approx\sqrt{\frac{2}{\pi A_1}}\textrm{e}^{-3x/2}\Big[C_1\cos{\Big(A_1\textrm{e}^{3x}-\frac{\pi}{4}\Big)}+C_2\sin{\Big(A_1\textrm{e}^{3x}-\frac{\pi}{4}\Big)}\Big],
\end{equation}
where 
\begin{equation}
A_1\equiv\frac{4}{\sqrt{3\kappa\hbar^2}}.
\end{equation}
Here $J_\nu(x)$ and $Y_\nu(x)$ are Bessel function of the first kind and second kind, respectively. It can be seen that the wave function vanishes when $a$ and $x$ go to infinity.

\subsection{Wheeler-DeWitt equation: factor ordering 2 \label{3.3}} 
From the WDW equation \eqref{wdwtotal}, we can as well derive a quantum Hamiltonian by choosing another factor ordering
\begin{equation}
\frac{\hat{p_b}^2}{b}=-\hbar^2\Big(\frac{1}{\sqrt{b}}\frac{\partial}{\partial b}\Big)\Big(\frac{1}{\sqrt{b}}\frac{\partial}{\partial b}\Big).
\label{pbb}
\end{equation}
Before proceeding further, we highlight that this quantization is based on the Laplace-Beltrami operator which is the Laplacian operator in minisuperspace \cite{KieferQG}. This operator depends on the number of degrees of freedom involved. For the case of a single degree of freedom, it can be written as in Eq.~\eqref{pbb}.

Under this factor ordering and after introducing a new variable $y\equiv(\sqrt{\lambda}b)^{3/2}$, in both cases (LR and LSBR) the WDW equation can be written as 
\begin{equation}
\Big(\frac{d^2}{dy^2}+\frac{64}{3\kappa\hbar^2}y^2\Big)\Psi(y)=0,
\label{diff2}
\end{equation}
when $a$ and $y$ approach infinity. The solution of the previous equation reads \cite{mathhandbook}
\begin{equation}
\Psi(y)=C_1\sqrt{y}J_{1/4}(A_1y^2)+C_2\sqrt{y}Y_{1/4}(A_1y^2),
\end{equation}
and when $y\rightarrow\infty$, therefore, \cite{mathhandbook}
\begin{equation}\label{approxPsiy}
\Psi(y)\approx\sqrt{\frac{2}{\pi A_1y}}\Big[C_1\cos{\Big(A_1y^2-\frac{3\pi}{8}\Big)}+C_2\sin{\Big(A_1y^2-\frac{3\pi}{8}\Big)}\Big].
\end{equation}
Consequently, the wave functions approach zero when $a$ goes to infinity. According to the DeWitt criterium for singularity avoidance \cite{DeWitt:1967yk}, the LR and LSBR is expected to be avoided independently of the factor orderings considered in this work.

\subsection{Expected values}

We have shown that the DeWitt criterium of singularity avoidance is fulfilled hinting that the universe would escape the LR and LSBR in the EiBI model once the quantum effects are important. We next estimate the expected value of the scale factor of the universe $a$ by estimating the expected value of $b$. The reason we have to resort to the expected value of $b$ rather than $a$ is that in the classical theory \cite{Delsate:2012ky} that we have quantized the dynamics is only endowed to the scale factor $b$. We remind at this regard that when approaching the LR and LSBR, $b\propto a\ln a$ and $b\propto a\sqrt{\ln a}$, respectively, at least within the classical framework. Therefore if the expected value of $b$, which we will denote as $\textbf{b}$, is finite, then we expect that the expected value of $a$; i.e. $\textbf{a}$ would be finite as well. Therefore, non of the cosmological and geometrical divergences present at the LR and LSBR would take place.   

We next present a rough estimation for an upper limit of $\textbf{b}$ for the two quantization procedures presented on the previous subsection.
\begin{itemize}
\item Factor ordering I:

The expected value of $b$ at late-time can be estimated as follows:
\begin{equation}
\textbf{b}=\int_{x_1}^{\infty}\Psi^{*}\left(x\right) \frac{e^{x}}{\sqrt{\lambda}}\Psi\left(x\right)dx,
\end{equation}
where $x_1$ is large enough to ensure the validity of the approximated potential in (\ref{difeq1}), i.e., $\delta\rightarrow 0$. In this limit, we can use the asymptotic behavior for the wave function c.f. Eq.~(\ref{approxPsix}). Then, it can be shown that the approximated value of \textbf{b} is bounded as
\begin{equation}
\int_{x_1}^{\infty}\Psi^{*}\left(x\right) \frac{e^{x}}{\sqrt{\lambda}}\Psi\left(x\right)dx<\frac{\left\vert C_1\right\vert^2+\left\vert C_2\right\vert^2}{\pi A_1\sqrt{\lambda}}e^{-2x_1}.
\end{equation}
Therefore, we can conclude that \textbf{b} has an upper finite limit. Consequently, the LR and LSBR are avoided.

\item Factor ordering II:

In this case the expected value of $b$ can be written as 
\begin{equation}\label{intb2}
\textbf{b}=\int_{y_1}^{\infty}\Psi^{*}\left(y\right) \frac{y^{\frac{2}{3}}}{\sqrt{\lambda}}\Psi\left(y\right)f\left(y\right)dy,
\end{equation}      
where $y_1$ is large enough to ensure the validity of the approximated  potential in (\ref{diff2}), i.e., $\eta\rightarrow0$. In addition, we have  introduced a phenomenological weight $f\left(y\right)$ such that the norm of the wave function is well defined and finite for large $y$ \cite{Barvinsky:1993jf,Kamenshchik:2012ij,Barvinsky:2013aya}. In fact, we could as well choose $f\left(y\right)=y^{-\alpha}$ with $2/3<\alpha$. After some simple algebra, we obtain
\begin{equation}
\textbf{b}<\frac{2\left(\left\vert C_1\right\vert^2+\left\vert C_2\right\vert^2\right)}{\pi A_1\sqrt{\lambda}}\int_{y_1}^{\infty}y^{-\frac{1}{3}}f\left(y\right).
\end{equation}
Consequently, we get
\begin{equation}
\textbf{b}<\frac{2\left(\left\vert C_1\right\vert^2+\left\vert C_2\right\vert^2\right)}{\pi A_1\sqrt{\lambda}\left(\alpha-2/3\right)}y_{1}^{\frac{2}{3}-\alpha}.
\end{equation}
Once again, we reach the conclusion that \textbf{b} is finite. Therefore, the LR and LSBR are avoided.
\end{itemize}
\section{Conclusions}
Singularities seem inevitable in most theories of gravity. It is therefore natural to ask whether by including quantum effects would the singularities be removed. In the case of the EiBI scenario, while the big bang singularity can be removed, the intrinsic phantom dark energy doomsday remains inevitable \cite{Bouhmadi-Lopez:2014jfa}. We solved the modified Wheeler-DeWitt equation of the EiBI model for a homogeneous and isotropic universe whose matter content corresponds to two kinds of perfect fluid. Those fluids within a classical universe would unavoidably induce  LR or LSBR. We show that within the quantum approach we invoked, the DeWitt criterion is fulfilled and it leads toward the potential avoidance of the LR and LSBR. Our conclusion appears unaffected by the choice of factor ordering.
\section*{Acknowledgments}

The work of IA was supported by a Santander-Totta fellowship ``Bolsas de Investiga\c{c}{\~a}o Faculdade de Ci{\^e}ncias (UBI) - Santander Totta''.
The work of MBL is supported by the Basque Foundation of Science Ikerbasque. She also wishes to acknowledge the partial support from the Basque government Grant No.~IT956-16 (Spain) and FONDOS FEDER under grant FIS2014-57956-P (Spanish government). This research work is supported partially by the Portuguese grand UID/MAT/00212/2013. CYC and PC are supported by Taiwan National Science Council under Project No. NSC 97-2112-M-002-026-MY3 and by Leung Center for Cosmology and Particle Astrophysics, National Taiwan University.

\end{document}